\documentclass{article}
\usepackage{eurosym}
\usepackage{amsmath,amssymb,amsthm}
\usepackage{graphicx}
\usepackage{subcaption}
\usepackage{verbatim}
\usepackage{bm}
\usepackage{color}
\usepackage{amsmath}
\usepackage{amsfonts}
\usepackage{amssymb}

\begin{document}

\title{Anticlinic order of long-range repulsive rod-like magnetic particles
in 2D}
\author{Xiaoyu Zheng$^{1}$ \thanks{%
Corresponding author, Email: xzheng3@kent.edu, ORCID 0000-0002-3787-7741},
Obeng Appiagyei Addai$^{1}$\thanks{
Email: oaddai@kent.edu, ORCID 0000-0003-0024-7989}, Peter Palffy-Muhoray$%
^{1,2}$\thanks{
Email: mpalffy@kent.edu, ORCID 0000-0002-9685-5489} \\
%EndAName
\emph{$^1$Department of Mathematical Sciences, Kent State University, OH, USA%
}\\
\emph{$^2$Advanced Materials and Liquid Crystal Institute, Kent State
University, OH, USA} }
\maketitle

\begin{abstract}
In the field of liquid crystals, it is well-known that rod-like molecules
interacting via long-range attractive interactions or short-range repulsive
potentials can exhibit orientational order. In this work, we are interested
in what would happen to systems of rod-like particles interacting via
long-range repulsive potential. In our model, each particle consists of a
number of point dipoles uniformly distributed along the particle length,
with all dipoles pointing along the $z-$ axis, so that the rod-like
particles repel each other when they lie in the $x-y$ plane. Dipoles from
different particles interact via an $r^{-3}$potential, where $r$ is the
distance between the dipoles. We have considered two model systems, each
with $N$ particles in a unit cell with periodic boundary conditions. In the
first, particle centers are fixed on a square or triangular lattice but they
are free to rotate. In the second, particles are free to translate as well
as in cells with variable shapes. Here they self-assemble to form
configurations where the stress tensors are isotropic. Our numerical results
show that at low temperatures the particles tend to form stripes with
alternating orientations, resembling herringbone patterns or the anticlinic
Sm-C$_{A}$ liquid crystal phase.
\end{abstract}

\section{Introduction}

It is well known that systems of rod-like molecules form the nematic phase
under certain circumstances. For thermotropic liquid crystals, the rod-like
molecules interact via induced dipole-dipole interactions. This long-range
attractive forces between molecules promotes alignment with each other. At
sufficiently low temperatures, the Helmholtz free energy of the system is
minimized when the molecules more or less align with each other, resulting
in long-range orientational order \cite{MS}. On the other hand, hard rod
systems, in which particles interact via a short-range hard-core
interparticle potential, form the nematic phase for certain aspect ratios
above certain packing fractions. The free energy depends solely on the
entropy, which is a measure of the orientation dependent free volume of the
system. At sufficiently high packing fractions, the rod-like particles form
orientational order, giving up the orientational entropy in order to gain
translational entropy \cite{Frenkel}.

It is interesting to ask whether a system of anisometric particles
interacting with a long-rang repulsive potential will spontaneously form
orientational order. To answer this question is the motivation for this work.

An immediate example of repulsive particles are discrete point charges with
the same sign interacting via pairwise $r^{-1}$potential, where $r$ is the
distance between charges. Wigner predicted that a gas of electrons at zero
temperature form a bcc lattice in 3D \cite{Wigner1934}. Two-dimensional
experiments showed that at low temperatures and high areal density, a sheet
of electrons crystalized into a triangular lattice in 2D \cite{Adams1979},
which is known as a Wigner crystal. Montgomery \cite{Montgomery1988} proved
that for particles in a 2D Bravais lattice interacting with pairwise $r^{-s}$
potential with fixed density, the triangular lattice is the unique minimizer
for any real $s>0$.

An example of a system of particles interacting via an $r^{-3}$ potential is
a system of discrete point electric or magnetic dipoles. In one realization,
the centers of the dipoles are confined to a 2D planar region, and the
direction of the dipoles are constrained to be along the normal to the plane
and all point in the same direction, so that they repel each other.
Experiments on 2D colloidal superparamagnetic particles interacting via
repulsive magnetic dipolar interactions clearly showed that the systems form
a crystal phase with a triangular lattice at low temperatures \cite%
{Maret1999, Keim2010}, consistent with the results in \cite{Montgomery1988}.

Here, we consider particles consisting of two or more dipoles in the $z-$%
direction, fixed on a long rod which lies in the $x-y$ plane.\ The rods
repel each other. A realistic object resembling this model is a thin slab of
magnetic material, magnetized through its thickness, lying in the plane. 
\textit{A collection of such slabs serves as a primitive model of the system
that we want to study in this work.} The pair potential may be separated
into an isotropic and an anisotropic part. The isotropic part accounts for
the interactions without regard to the shape, and the anisotropic part for
the interactions due to the shape or the anisotropic distribution of
dipoles. As discussed below, at low number densities, the isotropic
interactions dominate, so one expects that the particle centers form a
crystal structure similar to that of point charges or dipoles. As the number
density increases, the anisotropic repulsive interactions become
predominant, so that the particle positions and orientations become
correlated. Therefore at high densities, the low density crystal structure
may be destroyed or deformed and new ordered structures may appear. We are
interested in finding these new ordered structures and in understanding
their density dependence.

A system of rod-like particles with long-range repulsion has been studied in
3D \cite{Liu2015}. The authors confined charged rod-like colloids in a thin
wedged-shaped cell and showed that the position and orientational order of
the colloids depends on the interaction with the cell walls. In our work, we
study configurations either formed by the particles alone, without external
constraints, or where the orientations of the particles are not affected by
external constraints.

The paper is organized as follows. Section \ref{Sect_model} lays out the
mathematical model describing the interactions between the particles in our
system. Section \ref{Sect_num} discusses the model systems that we are
interested in and the numerical procedures which lead to the identification
of equilibrium states. Section \ref{Sect_results} presents the numerical
results on systems of particles on square and triangular lattices and on
particles in cells with variable shapes and we conclude in Section \ref%
{Sect_conclusion}.

\section{Mathematical Model}

\label{Sect_model}

In this work, we ignore thermal effects and the kinetic energy of particles.
We look for configurations which correspond to minima of the potential
energy. The total potential energy of the system of $N$ particles, assuming
pairwise interactions, is given by 
\begin{equation}
E=\frac{1}{2}\sum_{\substack{ i=1}}^{N}\sum_{j\neq i}^{N}w(\mathbf{q}_{i},%
\mathbf{q}_{j}),
\end{equation}%
where $\mathbf{q}_{i}=(\mathbf{r}_{i},\theta _{i})$ is the generalized
coordinate of particle $i$, with $\mathbf{r}_{i}=(x_{i},y_{i})$ the
coordinates of the center, and $\theta _{i}$ the angle formed by the long
axis of the particle with the $x-$axis. The system may be subject to
external constraints, for example, geometric confinements or periodic
boundary conditions.

We model each particle as a long rod with $n$ uniformly distributed point
dipoles along the particle length, where all the dipoles point along the $z-$%
direction. We set the width of each particle to be zero, as the width does
not play a role in our model, so the particle shape is a line segment. The
interpenetration of the particles or of the line segments, is prohibited.
One can also consider particles as being made up by two lines of charges of
finite length, with the line of positive charges above the plane, and the
line of negative charges beneath the plane. Our model of particles with
discrete dipoles is an approximation of a line of dipoles with finite
length, with dipoles pointing to the $z-$direction. With this in mind, the
interaction between two particles can be modeled by the sum of all pair
interactions between dipoles, excluding dipole pairs on the same particle,%
\begin{equation}
w(\mathbf{q}_{i},\mathbf{q}_{j})=\sum_{k=1}^{n}\sum_{l=1}^{n}w_{(i,k)(j,l)}.
\label{eqn_interaction_2}
\end{equation}%
Here we use the notation $w_{(i,k)(j,l)}$ to denote the interaction energy
of the $k^{th}$ dipole on $i^{th}$ particle and $l^{^{th}}$ dipole on $j^{th}
$particle. The interaction energy of two permanent magnetic point dipoles is
given by \cite{Griffiths1968} 
\begin{equation}
w_{ab}=-\frac{\mu _{0}m_{a}m_{b}}{4\pi r^{3}}(3(\mathbf{\hat{m}}_{a}\cdot 
\mathbf{\hat{r}})(\mathbf{\hat{m}}_{b}\cdot \mathbf{\hat{r}})-(\mathbf{\hat{m%
}}_{a}\cdot \mathbf{\hat{m}}_{b})=\frac{\mu _{0}m_{a}m_{b}}{4\pi r^{3}},
\end{equation}%
where $\mu _{0}$ is the permeability of free space, $\mathbf{m}_{a}=m_{a}%
\mathbf{\hat{z}}$ is the dipole moment and $\mathbf{r}=r\mathbf{\hat{r}}$ is
the vector from $\mathbf{m}_{b}$ to $\mathbf{m}_{a}$.

In the minimum energy configuration, the force and torque on each particle
are zero. The total force on each particle is the sum of forces on all
dipoles on the particle,%
\begin{equation}
\mathbf{F}_{i}=\sum_{k=1}^{n}\mathbf{F}_{i,k},  \label{eq_force}
\end{equation}%
where $\mathbf{F}_{i,k}$ is the force acting on the $k^{th}$ dipole of
particle $i$, with%
\begin{equation}
\mathbf{F}_{i,k}=\sum_{j\neq i}^{N}\sum_{l=1}^{n}\mathbf{F}_{(i,k)(j,l)},
\label{eq_force_2}
\end{equation}%
and $\mathbf{F}_{(i,k)(j,l)}$ is the force acting on the $k^{th}$ dipole of
particle $i$ due to the $l^{th}$ dipole of particle $j$. One can readily
calculate the force $\mathbf{F}_{ab}$ on $\mathbf{m}_{a}$ due to the
presence of $\mathbf{m}_{b}$ as the negative gradient of the pair potential, 
\begin{equation}
\mathbf{F}_{ab}=-\nabla w_{ab}=\frac{3\mu _{0}m_{a}m_{b}}{4\pi r^{4}}\mathbf{%
\hat{r}}.  \label{eq_force_3}
\end{equation}%
The torque on each particle with respect to the particle center is the sum
of torques on all the dipoles on the particle,%
\begin{equation}
{\boldsymbol{\tau }}_{i}=\sum_{k=1}^{n}(\mathbf{r}_{i,k}-\mathbf{r}%
_{i})\times \mathbf{F}_{i,k}.  \label{eq_torque}
\end{equation}

We define the number density $\rho =N/A$, where $A$ is the area of the
region containing the particles. \ There are three lengths in our system:
domain size, particle size, and mean interparticle distance $\sqrt{1/\rho }$%
. We define the particle length $l$ as the distance between the two farthest
dipoles on a particle. If we scale all the lengths by the particle length $l$%
, and scale the dipole moments by $m_{0}$, then the scaled pair potential
and force between two dipoles are 
\begin{eqnarray}
\bar{w}_{ab} &=&\frac{w}{\mu _{0}m_{0}^{2}/4\pi l^{3}}=\frac{1}{r_{c}^{3}},
\\
\mathbf{\bar{F}}_{ab} &=&\frac{\mathbf{F}_{ab}}{\mu _{0}m_{0}^{2}/4\pi l^{4}}%
=\frac{3}{r_{c}^{4}}\mathbf{\hat{r},}
\end{eqnarray}%
where $r_{c}$ is the normalized distance between two point dipoles.

\begin{figure}[htb]
\centering
a)\includegraphics[width=.4\linewidth]{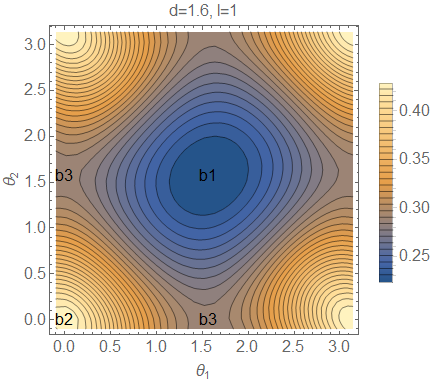} \qquad b)%
\includegraphics[width=.45\linewidth]{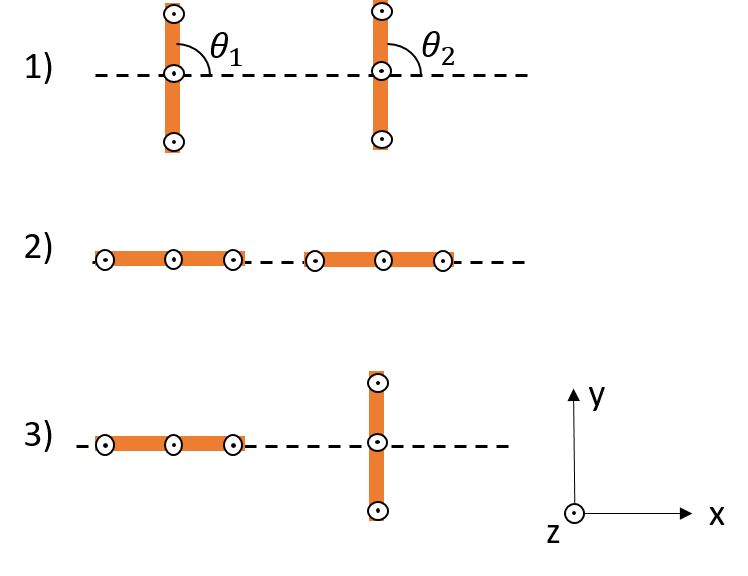}
\caption{(a) Representative energy contours of two particles separated on
the $x-$axis by a distance of $d=1.6$ as function of the angles $\theta_1$ and $\theta_2$. The number of dipoles on each particle is $n=21$. The minimum energy
state is attained when both angles are $\protect\pi/2$, corresponding to the
configuration in (b1). The maximum energy
state is attained when both angles are zero, corresponding to the configuration in (b2). If $\theta_1=0$, then the minimum energy state is attained when $\theta_2=\pi/2$, corresponding to the configuration in (b3). In the schematic of (b), only three dipoles are shown on
each particle to illustrate the directions of dipoles.}
\label{fig_contour2}
\end{figure}

We consider the elementary case of two particles centered along the $x-$
axis and separated by a distance $d$. The pair interaction energy, if $d\gg
1 $, can be written as 
\begin{equation}
\bar{w}(d,\theta _{1},\theta _{2})=\frac{n^{2}}{d^{3}}+\frac{(n+1)n^{2}}{%
16(n-1)d^{5}}[6+5(\cos 2\theta _{1}+\cos 2\theta _{2})]+O\left( \frac{1}{%
d^{7}}\right) ,
\end{equation}%
where the first term is an isotropic term in which only centers matter, and
the rest contain orientation dependent anisotropic terms. Considering only
the first two terms, the minimum occurs when $\theta _{1}=\theta _{2}=\pi /2$%
. Keeping more terms makes it difficult to determine where the minimizer is
analytically. Instead, we plot the energy contours in terms of the
orientations of the two particles. A typical plot is shown in Fig.~\ref%
{fig_contour2}(a) with the number of dipoles $n=21$ at a distance $d=1.6$.
The energy contour plots show that the minimum energy state occurs when $%
\theta _{1}=\theta _{2}=\pi /2$, for any $d>0$. That is, they are parallel
to each other and both are perpendicular to the line connecting the centers,
as shown in Fig.~\ref{fig_contour2}(b1). The energy attains its maximum when
they are both parallel to the line connecting the centers, as shown in Fig.~%
\ref{fig_contour2}(b2), and attains its saddle point when one aligns along
the $x-$direction and the other along the $y-$direction, as in Fig.~\ref%
{fig_contour2}(b3). If the separation distance between the centers is too
short, then the configurations in (b2) and/or (b3) in Fig.~\ref{fig_contour2}
cannot be attained as the particles will interpenetrate, but the minimum
energy state is still attained at the configuration in (b1), which is the
minimizer of the anisotropic part of the energy.

If we consider the one-dimensional problem where a number of such particles,
whose centers are constrained to be on an interval on the $x-$axis, with
periodic boundary conditions, then in the minimum energy state the centers
of the particles are uniformly distributed in the interval and all particles
all parallel to the $y-$axis. In two dimensions, this simple configuration
where all particles are aligned parallel to the $y-$axis is no longer be the
minimum energy state. What configuration will minimize the total energy in
this case? We try to answer this question in two model systems detailed
below.

\section{Model systems and numerical procedures}

\label{Sect_num}

We are interested in systems where the particles will form patterns in a 2D
plane. For this, we have considered two specific model systems: (a) systems
where the particle centers are fixed on a given lattice, and (b) systems in
which particles self-assemble into some recognizable configurations. In both
cases, we consider a finite system with periodic boundary conditions.

In the first model system, there is an underlying lattice $\mathbf{L}$,
where the particle centers are fixed at each lattice site, and the goal is
to find particle orientations which minimize the total potential energy. The
underlying lattice can be regarded as an external constraint, but such a
constraint will not directly influence the orientations of particles. In
this case, the energy of the system is given by%
\begin{equation}
E_{L}=\frac{1}{2}\sum_{i\neq j\in \mathbf{L}}w(\mathbf{q}_{i},\mathbf{q}%
_{j}).
\end{equation}%
Here the sum is over all lattice sites. Since analytic expressions for the
sum are unavailable, we solve the problem numerically. Our numerical
procedure is as described below.

We first partition an infinite lattice into identical unit cells, in the
shape of parallelograms, each containing a number of lattice sites. The unit
cell is then surrounded by 8 congruent image cells. If we denote the edges
of a unit cell by vectors $\mathbf{a}$ and $\mathbf{b}$, and denote the
matrix $H=[\mathbf{a},\mathbf{b}]$, then the area of the cell is given by $%
A=\det H$. The position of the center of particle $i$ can be expressed as a
linear combination of the two basis vectors as $\mathbf{r}_{i}=\xi _{i}%
\mathbf{a+}\eta _{i}\mathbf{b}$ $=H\mathbf{t}_{i}$, where $\mathbf{t}_{i}%
\mathbf{=(}\xi _{i},\eta _{i}\mathbf{)}^{\prime }$ is the relative
coordinate of the particle $i$ with $0<\xi _{i},\eta _{i}<1$. The centers of
particles in the image cells can be assigned as $\mathbf{r}_{i}+H\mathbf{n}$%
, where $\mathbf{n}=(n_{1},n_{2})^{\prime }$, with integer components
ranging from $-1$ to $1$. Our numerical simulation starts with a given
initial orientations of all particles on the lattice sites in the unit cell,
which is then copied to all image cells. We next calculate the torque on
each particle according to Eq.~(\ref{eq_torque}). We assume that the
contributions to the torque on a particle from dipoles far away have a
minimal effect comparing to that from the nearby dipoles, thus the torque on
each particle is only calculated from dipoles within a cutoff radius.
Specifically, for each dipole on a particle, all dipoles on other particles
within a radius $r_{0}$, including those in the image cells, will be
included in the torque calculation, and dipoles outside will have no
contribution. In this work, the radius $r_{0}$\thinspace\ is taken to be the
length of the short side of the unit cell. We approach the equilibrium using
overdamped dynamics; we advance the system by a timestep $\Delta t$ where
each particle rotates with angular velocity proportional to the torque
acting on it. Discretizing this, the angles are updated as 
\begin{equation}
\theta _{i}(t+\Delta t)=\theta _{i}(t)+\beta \tau _{i}\Delta t,\text{ }%
i=1,...,N,
\end{equation}%
where $\beta $ is proportional to the reciprocal of rotational viscosity and 
$N$ is the number of lattice sites, or equivalently the number of particles
in the unit cell. Once the orientations of all particles in the unit cell
have been updated and copied to all image cells, the new torque is evaluated
for each particle based on the updated configuration. This process is
repeated until the torques on all particles vanish or are within a tolerance
of $10^{-12}$.

In the second model system, we consider $N$ particles free to move in a unit
cell shaped as a parallelogram. Periodic boundary conditions are imposed as
before such that the unit cell is surrounded by 8 congruent image cells. The
simulation starts with given initial positions and orientations of the $N$
particles in the unit cell, which is then copied to all image cells. Next,
the force and torque exerted on each particle in the unit cell from
neighboring dipoles within a cutoff radius are calculated according to Eqs.~(%
\ref{eq_force})$-$(\ref{eq_torque}). Using discretized overdamped dynamics
as before, each particle will then translate and rotate with velocity and
angular velocity proportional to the force and torque, 
\begin{eqnarray}
\mathbf{r}_{i}(t+\Delta t) &=&\mathbf{r}_{i}(t)+\alpha \mathbf{F}_{i}\Delta
t, \\
\theta _{i}(t+\Delta t) &=&\theta _{i}(t)+\beta \tau _{i}\Delta t,
\end{eqnarray}%
where $\alpha $ and $\beta $ are proportional to the reciprocals of the
viscosity and angular viscosity, respectively. At each step, the
orientations and positions of the particles are copied to all image cells.
This process is repeated until the forces and torques on all particles in
the cell vanish. Our journey does not end here. Since the particles are not
fixed on lattice sites, forces acting on them can give rise to stresses on
the unit cell walls, which can change the shape of the unit cell at fixed
number density.

For a given configuration, the average stress tensor acting on the unit cell
can be calculated by \cite{Louwerse2006}%
\begin{equation}
{\boldsymbol{\sigma }}=\frac{1}{2A}\sum_{i,k}\sum_{j(\neq i),l}\mathbf{F}%
_{(i,k)(j,l)}(\mathbf{r}_{(i,k)}-\mathbf{r}_{(j,l)}),  \label{eq_stress}
\end{equation}%
where the outer sum indices are over the dipoles in the central unit cell,
and the inner sum indices are over the dipoles in central unit cell as well
as the image cells, excluding the dipole pairs on the same particle. We
assume that the unit cell deforms in response to stress similarly to a
volume conserving elastic body. If the stress tensor is not isotropic, the
shape of the unit cell will change \cite{Parrinello1980}. In our simulation,
since the stress is symmetric, we have kept the vector $\mathbf{a}=(a,0)$
parallel to the $x-$axis, and let $\mathbf{b}=(b_{x},b_{y})$ free to rotate.
Specifically, the length of $\mathbf{a}$ is updated by 
\begin{equation}
a_{x}^{\prime }=a_{x}+Wb_{y}({\boldsymbol{\sigma }}_{xx}-{\boldsymbol{\sigma 
}}_{yy}),  \label{eq_shape_1}
\end{equation}%
where $W$ is proportional to the reciprocal of elastic modulus, and 
\begin{equation}
b_{y}^{\prime }=\frac{A}{a_{x}^{\prime }},  \label{eq_shape_2}
\end{equation}%
such that the area of the parallelogram remains as a constant. The $x-$%
component of vector $\mathbf{b}$ is updated as%
\begin{equation}
b_{x}^{\prime }=b_{x}+Wa_{x}{\boldsymbol{\sigma }}_{xy}.  \label{eq_shape_3}
\end{equation}%
Our simulation is in clear contrast from the traditional constant pressure
simulation, where a pressure tensor is imposed; rather, it is a constant
volume (area, in our case), variable cell shape simulation. The equilibrium
configurations correspond to the cell shapes which admit isotropic stress
tensors.

The final goal is to find the specific shape of the parallelogram such that
the stress tensor is isotropic, at which point the total energy is minimized
over all possible parallelograms with constant area.

We remark that when we consider the dipole interactions within a cutoff
radius, it is not always meaningful to compare the energies of the systems
with different underlying lattices, or even for particles on the same
lattice but with different particle orientations. There are two main reasons
for this. First, the energy decays slowly as function of system size.
Second, the number of participating dipoles is not always the same for
different configurations, and the energy calculation is largely dependent on
the number of pairs of dipoles involved. We therefore rely on the
calculations of torques and forces which vanish in equilibrium.

In general, there are multiple local energy minima, representing the rugged
energy landscape. We have carefully selected the simulation parameters so
that we can find the global minimum energy states and avoid frustrations by
only locally favored configurations.

\section{Results}

\label{Sect_results}

In this section, we present numerical findings of the two model systems
mentioned in Sect.\ \ref{Sect_num}. In the first model system, particle
centers are fixed at the sites of two simple lattices: a square lattice and
triangular (or hexagonal) lattice, and the particles are free to rotate
about their centers. In the second model system, particles are free to
translate as well as rotate in a cell with variable shape and constant area.

\subsection{Particles on square lattices}

\label{Sect_Square}

We first consider the case when the centers of particles are fixed on sites
of a square lattice. In this numerical experiment, the lattice size is
varied to account for different number densities $\rho $. We have considered
the cases where the total number of particles $N=16,36,64,100$ on a square
lattice in a unit cell. The results\ of the angles $\theta _{i}$ for
different values of $N$ agree to within $10^{-4}$ for all number densities.
Therefore, here we only present the representative results for $N=36$ with $%
n=9,21$ and $41$ dipoles on each particle.

\begin{figure}[tbh]
\centering
(a) \includegraphics[width =0.5\linewidth]{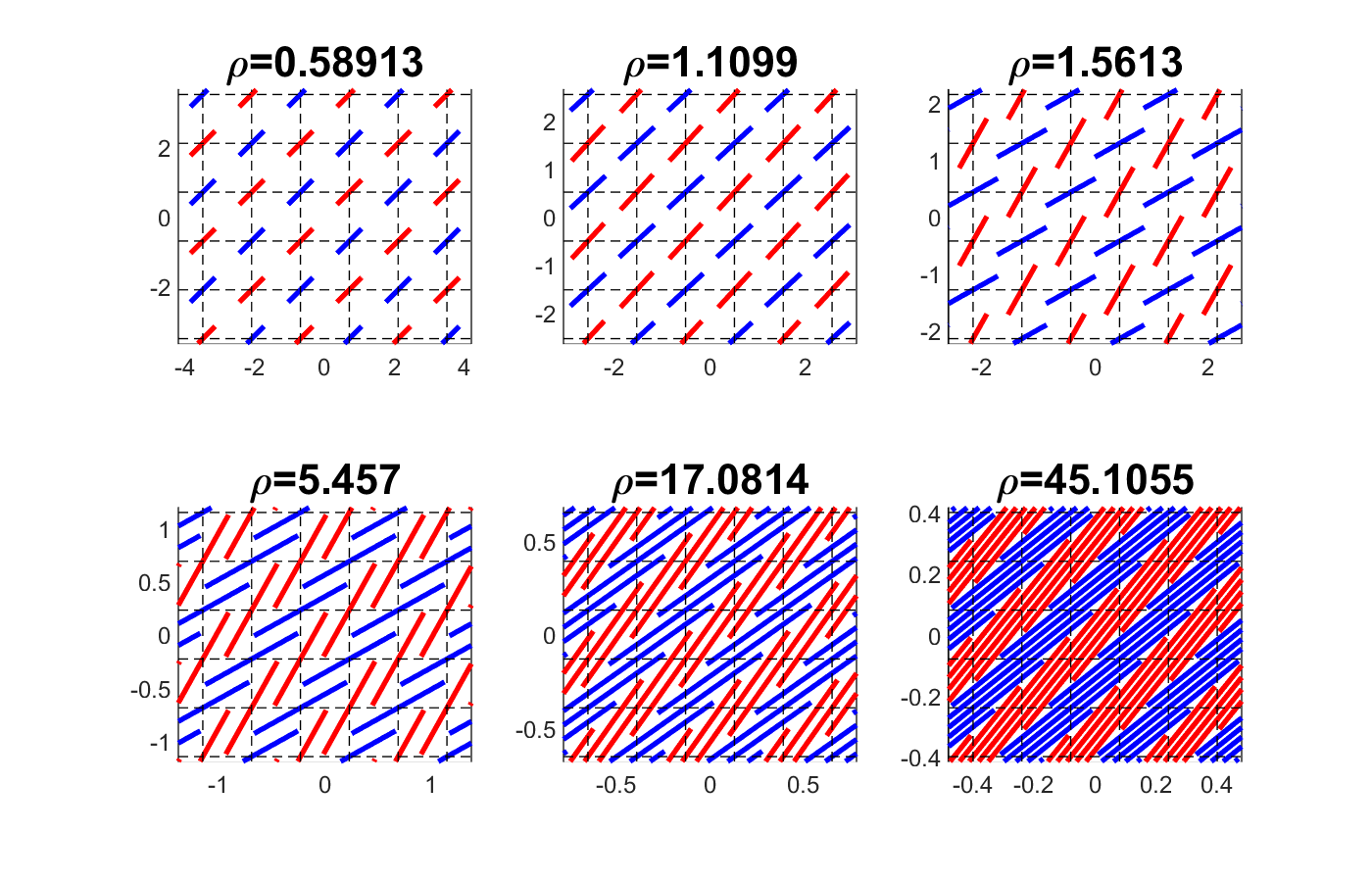} \quad (b) %
\includegraphics[width=0.4\linewidth]{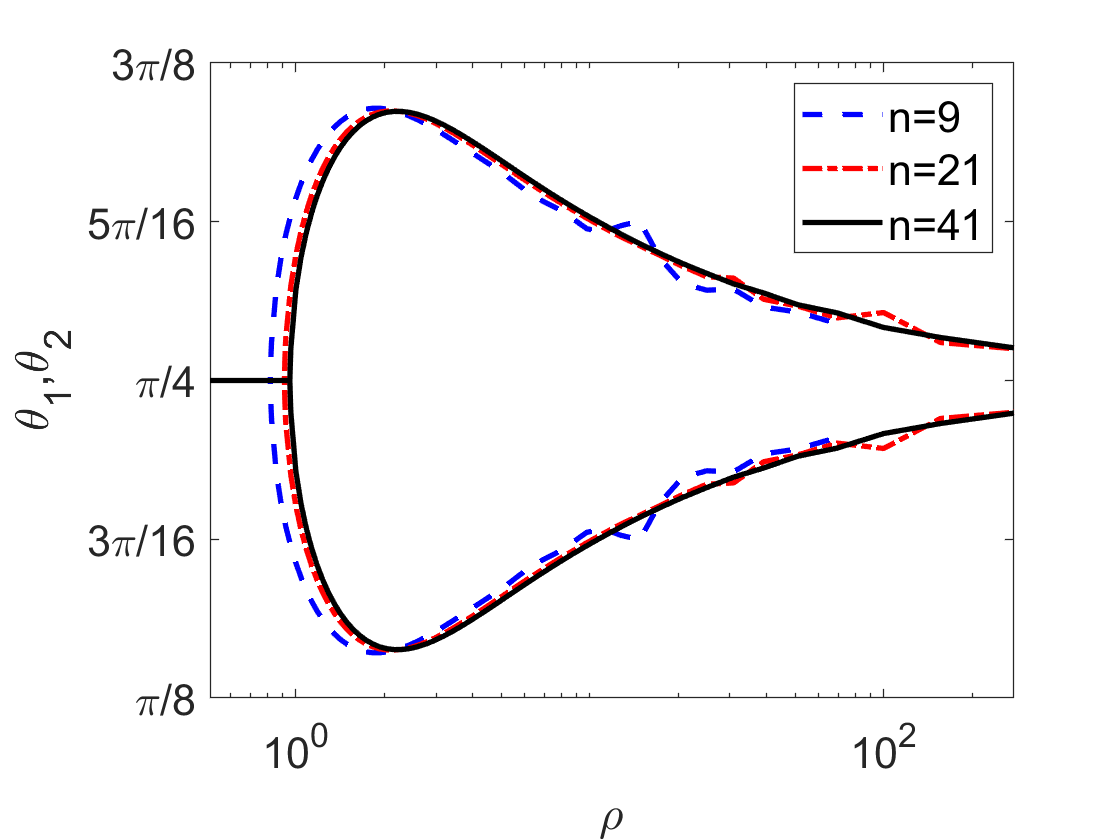}
\caption{ (a) Equilibrium configurations of particles on a square lattice
at 6 representative number densities. Each particle has 41 uniformly
distributed dipoles along its length. (b) The angles of particles in two
adjacent stripes on the square lattices as function of the number density $%
\protect\rho $.  }
\label{fig_squareconf}
\end{figure}

Figure \ref{fig_squareconf}(a) shows some representative equilibrium
configurations of particles on square lattices at several number densities,
with $n=41$ dipoles on each particle. For small values of $\rho $, that is,
when the particles are far apart, they align uniformly along one diagonal of
the lattice. In this configuration, the forces and torques from the four
nearest neighbor particles as well as from the four neighbors at the nearest
diagonal sites cancel exactly. As $\rho $ increases above a critical number
density, the configuration with the uniform alignment is no longer stable,
as dipoles at the two nearest diagonal sites along the orientation of
particles are so close that along the diagonal neighbor interactions
dominate, and the configuration becomes unstable. To lower the energy, the
particles on the neighboring sites rotate in opposite directions by the same
amount, which results in the configurations that all particles along any
diagonal, which is parallel to southwest-northeast direction, form a stripe.
Furthermore, in each stripe the particles are aligned in the same direction,
and the orientations alternate from stripe to stripe. The average
orientation of all particles is parallel to the direction of the stripe.
This configuration is also known as herringbone pattern.

Figure \ref{fig_squareconf}(b) shows the angles of particles in two adjacent
stripes as a function of $\rho $. Different curves correspond to the results
from $n=9,21,41$, respectively. The critical number densities where there is
a configurational transition from uniform alignment to the herringbone
pattern occurs at $\rho _{c}\approx 0.8175,0.9139,0.9536$ for $n=9,21,41$,
respectively. Larger values of $n$ shift the transition towards $\rho _{c}=1$%
. At high densities, the particles are getting close and their neighbors can
no longer be regarded as a continuous uniform body for small values of $n$.
As a result, the angle $\theta $ as function of $\rho $ for $n=9$ does not
follow a smooth curve, but instead oscillates about a smooth curve when $%
\rho \gtrapprox 10$. The curves of $\theta $ in terms of $\rho $ for $%
n=21,41 $ also oscillate but the oscillating behavior is deferred to higher
densities, where the numbers of dipoles per particle are not large enough to
be regarded as continuously distributed. From the simplest geometric view,
we anticipate that the orientation of particles becomes more and more
parallel to the stripe direction as $\rho $ increases for particles with
continuously distributed dipoles along their lengths. 

\subsection{Particles on triangular lattices}

\label{Sec_triangle}

Next, we fix the centers of particles on a triangular lattice. The numerical
procedure and simulation parameters are exactly the same as those on the
square lattices.

\begin{figure}[tbh]
\centering
(a)\includegraphics[width =0.5\linewidth]{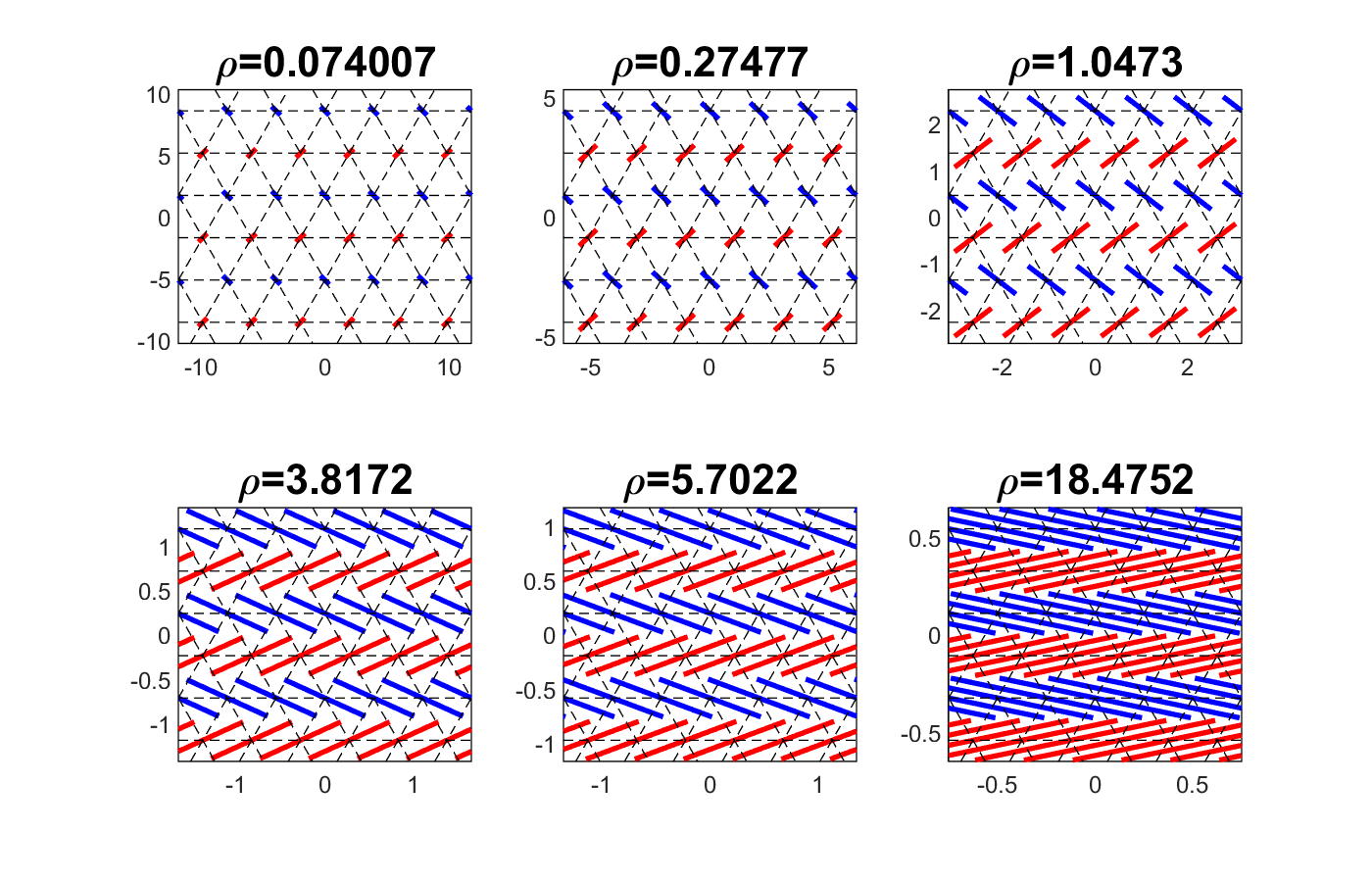}  (b)%
\includegraphics[width=.4\linewidth]{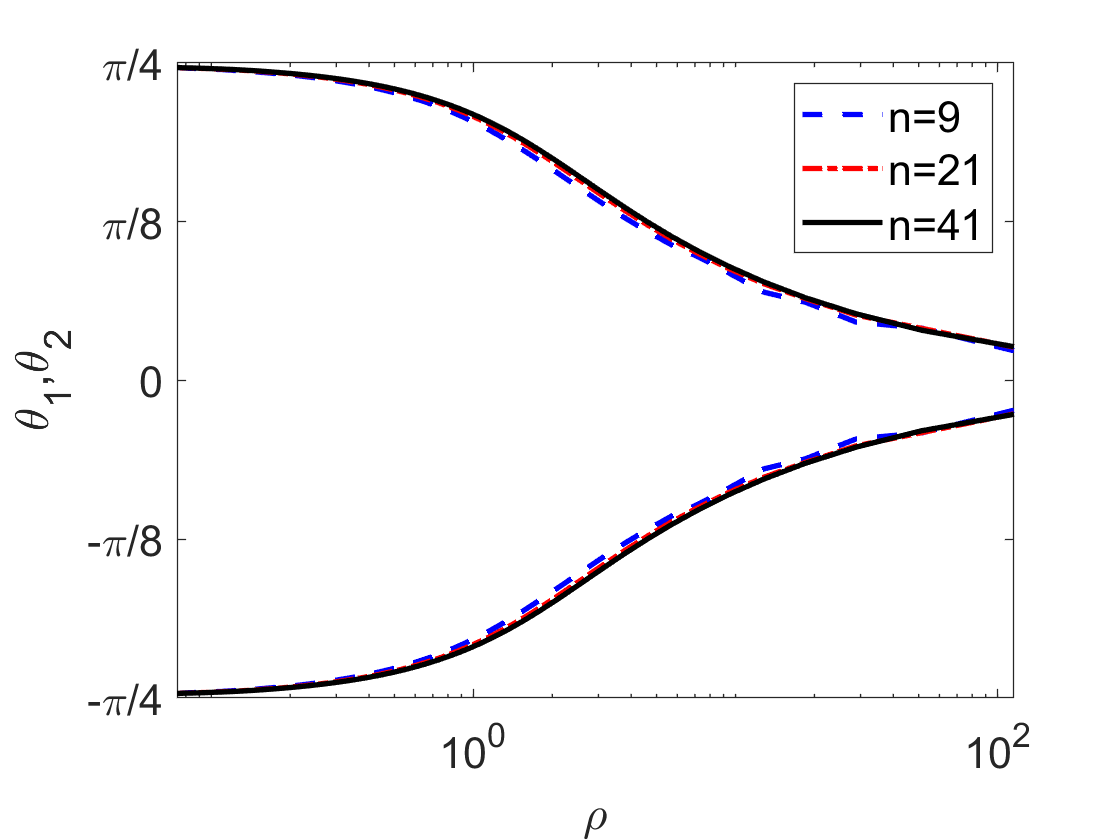}
\caption{(a) Equilibrium configurations of particles on a triangular
lattice at 6 representative number densities. Each particle has 41 uniformly
distributed dipoles. (b) The angles of the particles in two adjacent stripes
on the triangular lattices as function of the number density $\protect\rho 
$. }
\label{fig_TriConfig}
\end{figure}

Figure \ref{fig_TriConfig}(a) shows representative equilibrium
configurations of particles on triangular lattices at several number
densities. The orientations of the particles always form horizontal (or
along any grid lines due to the three-fold symmetry) stripes on a triangular
lattice, with alternating orientations from stripe to stripe. On average,
all particles point along the stripe direction. Figure \ref{fig_TriConfig}%
(b) shows the angles of the particles in two adjacent stripes as a function
of $\rho $. Each curve corresponds to $n=9,21,41$, respectively. Large
numbers of dipoles per particle make the angle between the particles in
adjacent stripes slightly larger. As $\rho \rightarrow 0$, the particle
orientations saturate at $\theta _{1}=-\theta _{2}=\pi /4$. As $\rho $
increases, the particles will tend to align with the stripe direction.

The results from the square and triangular lattices share one feature: the
particles at high densities form stripes with alternating orientations. 

\subsection{Free particles in a cell with variable shape}

One can find the equilibrium configurations of particles with different
underlying lattices and number densities following the same procedure.
However, we are more interested in the question of whether particles would
spontaneously organize themselves and form a lattice? If so, what lattices
would be most likely to form at different number densities?

As we have indicated earlier, at a given number density, the unit cell
containing the particles will adopt a shape such that the stress tensor of
the corresponding equilibrium configuration is isotropic. In fact, the
square and triangular lattices do not give rise to isotropic stress tensors.
In the square lattice case, the normal stress components are the same, but
the shear component is nonzero. In the triangular lattice case, the normal
stress components differ.

We have carried out simulations of $N$ particles in a unit cell under the
periodic boundary conditions, with different initial conditions and numbers
of particles. Since our goal is to identify the lattice formed by particles,
we redefine the unit cell to be nearly square during the course of the
simulation in order to maximize the cutoff length.

Preliminary results suggest that the particles tend to arrange themselves in
a centered rectangular lattice. Furthermore, the particles at the
rectangular vertices of the lattice have the same orientation, and the ones
in the center of the lattice have opposite orientations such that the
average angle is close to $0$. With this in mind, we have rerun the
simulations with centers of particles at a centered rectangular lattice, and
assign an angle $\theta $ to particles on the rectangular vertices and
assign the angle $-\theta $ to particles located at the center of the
lattice. We then find the angle $\theta $ and the aspect ratio of the
lattice as a function of $\rho $, where we define the aspect ratio of the
centered rectangular lattice as the ratio of short side of the rectangle and
the long side.

\begin{figure}[tbh]
\centering
(a)\includegraphics[width=1.\linewidth]{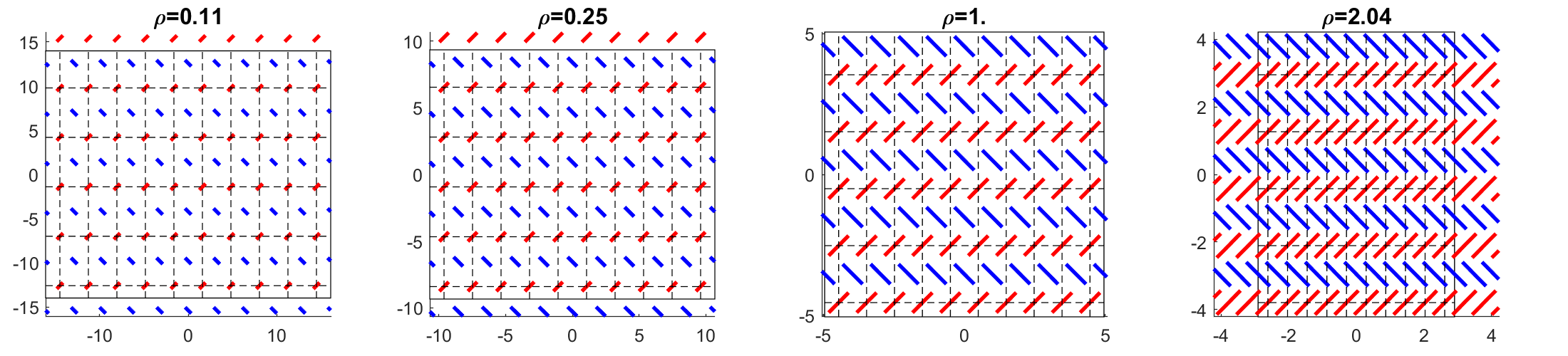} (b)%
\includegraphics[width=.45\linewidth]{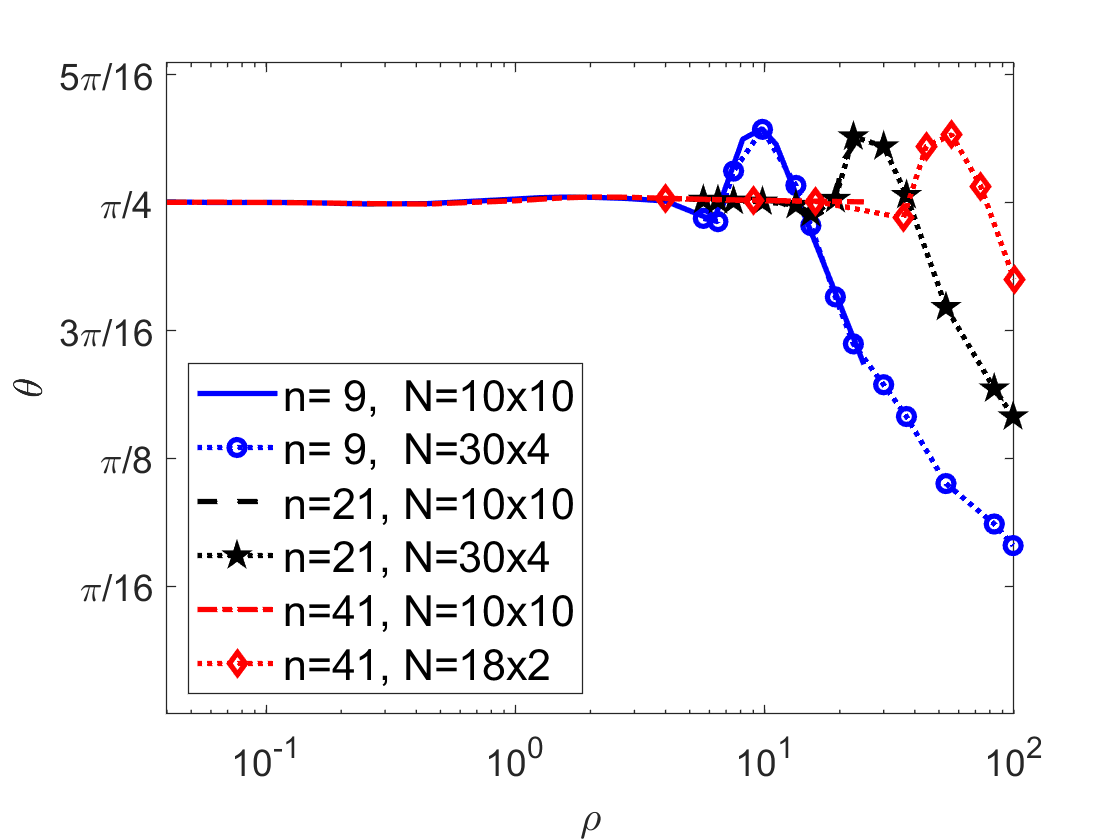}\qquad (c)%
\includegraphics[width=.45\linewidth]{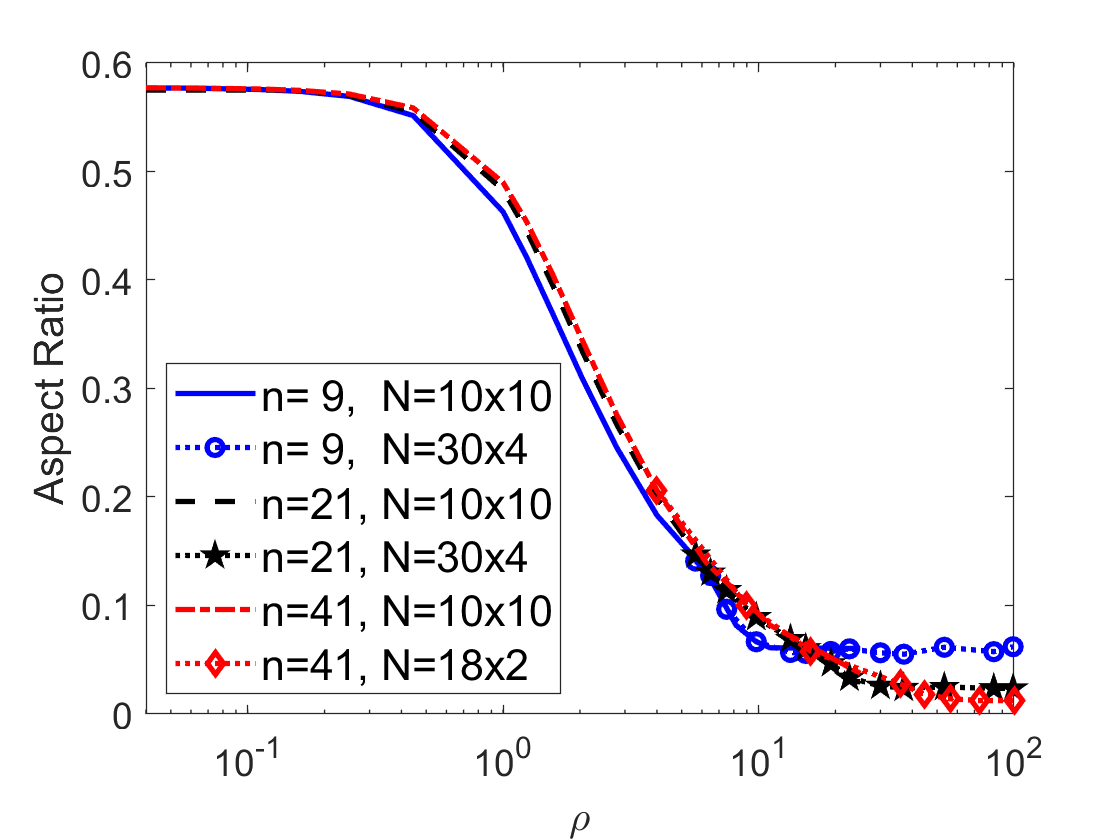}
\caption{(Color online)(a) Equilibrium configurations of 100 free particles
in a unit cell at 4 representative number densities. The unit cell is
bounded by thin solid lines, and the centered rectangular lattice is
indicated by the dashed lines. Each particle has 41 dipoles uniformly
distributed along its length. The red particles are on the vertices of the
centered rectangular lattice, and the blue particles are at the center of
the lattice. (b) The angle of the particles on the vertices of the centered
retangular lattice as function of the number density $\rho$. (c) The aspect ratio
of the centered rectangular lattice as function of the number density $\rho$.}
\label{fig_conf_movable_final}
\end{figure}

Our final results are summarized in Fig.~\ref{fig_conf_movable_final}. For
small number densities, we have considered $10\times 10$ particles in a unit
cell. Several representative equilibrium configurations are shown in Fig.~%
\ref{fig_conf_movable_final}(a) with different number densities. The unit
cell is bounded by thin solid lines, and the centered rectangular lattices
are indicated by the dashed lines. We have used different numbers of dipoles 
$n=9,21,$ and $41$ on each particle as before. In Fig.~\ref%
{fig_conf_movable_final}(a) we show the results from $n=41$. At high
densities, since the aspect ratio of the lattice is very small, the unit
cell is far from a square, so we considered more particles in the horizontal
direction and less in the vertical direction, e.g., $30\times 4$ or $%
18\times 2$, whose results are plotted in Fig.~\ref{fig_conf_movable_final}%
(b,c) at high number densities.

At low densities, the particles are far apart, so they can be regarded as
point particles and the shape anisotropy does not play a significant role in
determining the position of centers, and the lattice formed by the particle
centers turns out to be very close to a triangular lattice. If we regard the
triangular lattice as a special case of a centered rectangular lattice, then
its aspect ratio corresponds to $1/\sqrt{3}\approx 0.577$. This is
consistent with the result that repulsive point particles, interacting via
the $r^{-s},s>0$ potential, form a triangular lattice in 2D \cite%
{Montgomery1988}. As shown in Fig.~\ref{fig_conf_movable_final}(a), the
particles form horizontal stripes, similar to that in triangular lattices.
The angles of particles in each stripe is close to $\pm \pi /4$ in the
dilute limit, which agrees with the results from Sect.~\ref{Sec_triangle}.
As $\rho $ increases, the particles get closer. If we assume an isotropic
shrinkage of the unit cell, it turns out that the vertical component of the
stress tensor becomes larger than the horizontal component. In fact, if we
only consider the contribution to the stress tensor from interactions of
particles from the same stripe, not only does the horizontal component
increase, but the vertical component also increases although by a smaller
amount. If we consider the contribution to the stress tensor from particles
from different stripes, the increase of the vertical component is much
larger than that of the horizontal component. As a result, the stress on the
upper and lower boundaries is larger than on the left and right boundaries,
thus the unit cell grows taller and thinner, and the aspect ratio of the
lattice decreases. Up to $\rho \approx 4$, the angles of particles on
adjacent horizontal stripes are very close to $\pm \pi /4$. This result is
not very sensitive to the values of $n$, and the aspect ratio of the
centered rectangular lattice is slightly larger for larger values of $n$, as
shown in Fig.~\ref{fig_conf_movable_final}(b,c).

As $\rho $ increases a little further, the angle $\theta $ first dips down
slightly and then increases. As $\rho $ gets extremely high to about $\rho
\approx 10$, the angle $\theta $ starts to decrease while the aspect ratio
of the lattice remains almost constant. We attribute this nonlinear behavior
of $\theta $ as function of $\rho $ to the finite number of dipoles
considered. As we increase $n$ to $21$ and $41$, the nonlinear behavior of $%
\theta $ persists, but is deferred to higher number densities. This is
similar to what we have observed in Sect.~\ref{Sect_Square}. We have run
simulations and confirmed that at a given density, as we increase $n$, the
angle $\theta $ approaches the proximity of $\pi /4$. A compelling analytic
argument for this magic angle has yet to be found.

There are also equilibrium configurations in which particle centers form a
centered rectangular lattice and all particles orient uniformly along one
edge of the rectangle. However, those configurations represent unstable
equilibria.

\section{Discussions and Conclusions}

\label{Sect_conclusion}

In this work, we have studied the equilibrium configurations of systems of
rod-like particles interacting via long-range repulsive interactions in a
two dimensional plane. We model those particles as line segments with
uniformly distributed discrete point dipoles, with all dipoles pointing
along the $z$-direction. We confine our study to dipoles interacting via an $%
r^{-3}$ potential. We anticipate that the results will be qualitatively the
same if the pairwise potential is $r^{-s}$ with any $s>0$.

We have considered two model systems. In the first, we study the equilibrium
orientations of particles with centers confined to a square or a triangular
lattice. In the second, we study how particles self-assemble when they are
free to translate and rotate in a cell with variable shape. In both model
systems, periodic boundary conditions are imposed on a central unit cell
containing $N$ particles. We have chosen simulation parameters to prevent
the formation of defects.

When the centers are fixed on a square lattice, particles will uniformly
align along one of the diagonals in the dilute limit. As number density
increases, there is a second order transition from the uniform alignment to
the herringbone pattern where particles form stripes along the diagonals
with alternating orientations from stripe to stripe. When the centers are
fixed on a triangular lattice, the particles form the herringbone patterns
for all number densities. When the particles are free to translate and
rotate in a cell with variable shape, the particles will self-assemble to
form a centered rectangular lattice, where the angle of the particles on the
rectangular vertices of the lattice is close to $\pi /4$ and the angle of
the particle at the center of the lattice is close to $-\pi /4$. The lattice
is close to a triangular lattice at dilute concentrations, and the aspect
ratio of the lattice decreases as number density increases. Again, the
particles form a herringbone structure with alternating orientations in
adjacent stripes.

The herringbone structure is a key feature of the antiferroelectric smectic $%
C_{A}$ phase formed by smectic liquid crystals consisting of elongated
chiral molecules \cite{Cepic2010}. Systems of molecules with zigzag-shaped
or bent core architectures can also exhibit the anticlinic phases \cite%
{Goodby2015}. In one of the molecular models \cite{Osipov2000}, it has been
shown that conventional dispersion and steric intermolecular interactions
cannot stabilize the Sm-C$_{A}$ phase, rather the orientational correlations
between transverse molecular dipoles, when they are located in adjacent
smectic layers, are responsible for stabilizing the Sm-C$_{A}$ phase. In 
\cite{Glaser2002}, the authors proposed that the presence of molecular-scale
fluctuations of the layer interface between the smectic layers provide an
entropic mechanism for exhibiting synclinic Sm-C order rather than the
anticlinic Sm-C$_{A}$ order. For materials exhibiting anticlinic order, the
interface fluctuations might be suppressed due to their bent molecular
conformation. To our knowledge, our system is the first system where solely
long-range repulsive interactions are responsible for the formation of
anticlinic order. Dipoles are as far as possible from each other in this
anticlinic configuration.

If there is an external geometric constraint, for example, a finite number
of particles are contained in a rectangular box, then the arrangement of
particles will be very sensitive to the shape of the box and the number of
particles. The equilibrium configuration will also depend on the
interactions between the particles and the boundary walls.

In this study, thermal effects have been ignored. If thermal effects are
included, at sufficiently high temperatures, the crystal structure formed by
the centers of particles will melt, and orientational order will disappear.
Other phases may possibly emerge. We defer this to a future study.

\section*{Conflicts of interest}

There are no conflicts of interest to declare.

\section*{Acknowledgment}

This work was inspired by a beautiful experiment made by Dr.~Mykhailo Pevnyi
when he was a graduate student at the liquid crystal institute at Kent state
University. This work was supported by the Office of Naval Research through
the MURI on Photomechanical Material Systems (ONR N00014-18-1-2624) and Air Force Research Laboratory through STTR
grant: Electronically Dimmable Eye Protection Devices ( FA8649-20-C-0011).

\end{document}